# Measuring the Complexity of Urban Form and Design

Geoff Boeing
Northeastern University
g.boeing@northeastern.edu

October 2018

Abstract

Complex systems have become a popular lens for analyzing cities and complexity theory has many implications for urban performance and resilience. This paper develops a typology of measures and indicators for assessing the physical complexity of the built environment at the scale of urban design. It extends quantitative measures from city planning, network science, ecosystems studies, fractal geometry, statistical physics, and information theory to the analysis of urban form and qualitative human experience. Metrics at multiple scales are scattered throughout diverse bodies of literature and have useful applications in analyzing the adaptive complexity that both evolves and results from local design processes. In turn, they enable urban designers to assess resilience, adaptability, connectedness, and livability with an advanced toolkit. The typology developed here applies to empirical research of various neighborhood types and design standards. It includes temporal, visual, spatial, scaling, and connectivity measures of the urban form. Today, prominent urban design movements openly embrace complexity but must move beyond inspiration and metaphor to formalize what "complexity" is and how we can use it to assess both the world as-is as well as proposals for how it could be instead.

# Introduction

Complex systems are large nonlinear systems of interacting components that can produce "emergent" phenomena and allow structure to self-organize. Nearly all real-world systems are inherently nonlinear and their emergent features pervade our world (Barabási 2007; Wells 2014; Turnbull et al. 2018). The term *complexity* refers to the higher-order phenomena arising from a system's many connected, interacting subcomponents and describes both dynamics (i.e., processes) and structure (i.e., patterns and configurations) (Batty 2005). *Emergence* refers to the appearance of these characteristics over time as the components self-organize (Goldstein 1999). Human societies and cities are examples of large, complex systems (de Roo and Rauws 2012). Principles of complexity have been applied in urban planning and design from collaborative rationality, to cellular automata and agent-based models, to the design of livable neighborhoods (Innes and Booher 1999; 2000; Batty 2005; Sanders 2008; de Roo 2010). Complexity problematizes rationality and certainty – due to emergence's ability to surprise – providing a frame to theorize prediction's limitations and approach "wicked problems," incrementalism, and collaborative place-making (Innes and Booher 2010; Yamu et al. 2016; Rauws 2017; Skrimizea et al. 2018).

A city's planned and emergent structure, connectedness, accessibility, stability, resilience, and robustness involve urban design and bridge qualitative theories of cities and quantitative studies of physical form, design, and transportation (Portugali 2006). Urban designers often discuss urban form and design projects in terms of complexity (e.g., Duany, Plater-Zyberk & Co. 2001; Macdonald 2002; Talen 2003; Alexander 2003; Sanders 2008; Congress for the New Urbanism 2015). These discussions borrow salient concepts of complexity theory, but often loosely or metaphorically rather than scientifically, making it difficult to assess claims and outcomes (Chettiparamb 2006; cf. Marshall 2012a; Dovey and Pafka 2016; Rauber and Krafta 2018). Nevertheless, various formulations of complexity have long been regarded as critical to urban design. It contributes to lively, enjoyable, walkable, healthy, and vital neighborhoods (Jacobs 1961; Macdonald 2002; Carlson et al. 2012; McGreevy and Wilson 2016). It implies resilience, robustness, connectivity, and access – playing into wider debates about sustainability and resource efficiency (Peter and Swilling 2014; Pugh 2014; Wells 2014). Complexity in urban design can be emancipatory, re-centering top-down practices as liberating bottom-up collaborations that restructure the built environment to improve equity, spatial justice, adaptiveness, and social contact and exchange (Byrne 2003; cf. Pettigrew and Tropp 2006).

Research in multiple literature streams has considered cities, ecosystems, and other physical phenomena in terms of systemic complexity. Urban design can build on these foundations to formalize, clarify, and assess values and claims about physical complexity in the built environment. However, no synthesis exists in the urban design literature to organize and link these bodies of scholarship. This paper aims to address this gap. First, it unpacks complexity's relevance to urban design. Then it reviews temporal, visual, spatial, scaling, and connectivity measures of complexity from various disciplines, exploring their relevance to urban form character and urban design outcomes. Finally, it collates these measures into a typology to formalize and assess design claims, project results, and the urban form, before concluding with implications for practice and scholarship.

## Background

### Complexity and Cities

Belying the simple definition of complexity provided in the introduction, disagreements exist about the domain, interpretation, and implications of complexity theory. Shiner et al. (1999) offer a useful abstract distillation of these debates and interpretations in terms of order and disorder, illustrated by Figure 1's broad categories of complexity. Category I is positively correlated with disorder and includes algorithmic complexity and most measures of entropy. Here, complexity is highest when objects are scrambled-up with the greatest variety and diversity. Category II is a convex function of disorder, peaking at some midpoint between order and disorder. This balances variety and structure, conforming to traditional definitions of complex adaptive systems (e.g., Gershenson and Fernandez 2012). Category III takes complexity to be related more purely to order or structure alone, favoring self-organization and emergence in which structure emerges from disorder. This paper uses these interpretative categories to consider measures of complexity in urban design (depending on context and character), focusing on the second: the Jacobsian balance between structure and variety. While these categories are mutually incompatible (e.g., complexity cannot be simultaneously high *and* low when disorder is high), they reflect aspects of complexity to consider and assess.

How does the experience of – or preference for – complexity vary from person to person and culture to culture? We may have some intuitive sense of the complexity of a place simply by observing it or moving through it, but how might this be formalized? Cities are complex human ecosystems: population, density, employment, wealth, traffic volume, etc. can be (potentially) identified and (potentially) calculated at various scales to describe the system's evolving state. We can explore its dynamics with differential equations, regression models, machine learning algorithms, cellular automata, or agent-based models. But systems dynamics, stock/flow modeling, and other such measures of a world "becoming" are less useful for the characterization and analysis of urban structure and physical patterns "as exists" or "as proposed" (cf. de Roo 2010).

For the latter, we study built environments as physical structures produced by human behavior. This emphasizes spatial form rather than dynamical processes. Through co-evolution, humans both shape their cities and neighborhoods and are in turn shaped by them. The resulting physical patterns compose the urban form and can be studied in terms of network character, fractal structure, diversity (of various sorts), and entropy. At higher levels of abstraction, we can analyze the resilience, robustness, and adaptive capacity of urban complex systems and how they respond to perturbation given their spatial patterns, structure, connectedness, and efficiency. This provides a critical link from measuring complexity to the goals of urban decision-making, design, and planning interventions (de Roo and Rauws 2012; Yamu et al. 2016; Rauws 2017).

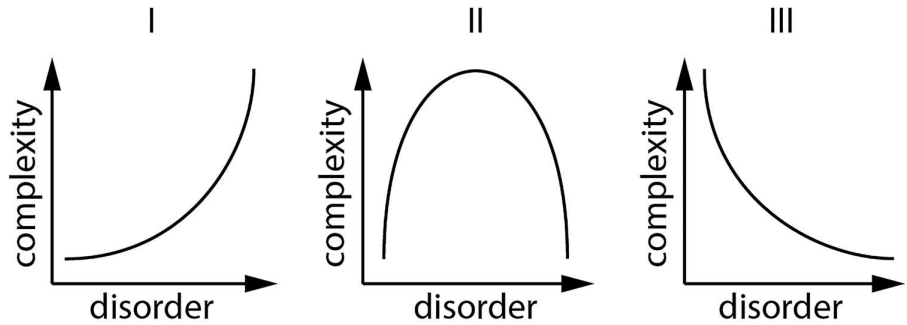

**Figure 1.** Three interpretations of complexity as a function of disorder. Category I increases with disorder. Category II peaks at a midpoint between order and disorder. Category III decreases as disorder increases. Image source: author; adapted from Shiner et al. (1999).

| Dimension | Description | Examples |
| --- | --- | --- |
| *temporal* | how processes and behavior change over time; unpredictability of human behavior and city futures | traffic jams<br>crowd behavior/dynamics<br>economic booms and busts<br>population growth and decline |
| *visual* | human perception of built environment's visual coherence, scale, interest, order, legibility, and detail | sense of enclosure<br>unity in variety<br>building façades and signage<br>human activity/vitality<br>sunlight patterns<br>tree canopy |
| *spatial* | land patterns and grain, particularly in terms of diversity | mixed land uses<br>racial/class integration/segregation<br>block sizes and shapes<br>economic agglomeration/clustering<br>spatial distributions of urban form elements |
| *scaling* | similarity of structure across multiple scales; fractal patterns | city area-perimeter allometrics<br>fractal urban form<br>surface textures<br>buildings of all sizes<br>streets of all sizes |
| *connectivity* | cities' and citizens' network organization, connectedness, circulation | communication and exchange<br>human travel patterns<br>destination accessibility<br>street connectivity and permeability<br>intersection types and density |

**Table 1.** Key dimensions of complexity in urban design.

**Urban Design and Physical Complexity**

Urban designers shape the public realm at the intersection of architecture and city planning (Moudon 1992; Cuthbert 2007; Biddulph 2012). Urban design includes centralized top-down acts, informal bottom-up acts, and everything in between. Its history reflects normative stances shifting through eras of classical formalism, romantic organicism, modernist simplifications, and post-Jane Jacobs gestures toward "organized complexity" (Barnett 2011). Jacobs's theories of complexity and bottom-up urbanism have been embraced by complex systems scholars studying cities – particularly from the physical sciences (Batty 2005; Bettencourt 2014; Batty and Marshall 2016; cf. O'Sullivan and Manson 2015).

In the 20$^{th}$ century, Corbusier and fellow modernists sought to eradicate the physical complexity of traditional cities, setting the stage for automobile dependency and single-use functional zoning (Hall 1996). Scott (1998) critiques modernist urban design by contrasting Corbusier's top-down, simplified, rational, polished, utopian cities with bottom-up, organically built, messy everyday urbanisms dependent on localized tacit knowledge (cf. Jacobs 1961; Holston 1989). Modernist designers confused geometric visual order for well-functioning, sustainable social order in the built environment (Roy 2005; Sussman and Hollander 2015). Scholars following in Jane Jacobs's wake have argued that simplified single-purpose urban design destroys functional capacity and synergy. Over-simplified interventions cut into the living tissue of urban complex systems, killing vital social processes: while healthy complex adaptive systems are resilient to perturbation, their resilience and adaptability can be compromised by too many simplifying interventions (Marshall 2012b).

How does urban design fit into the spectrum of bottom-up complexity versus top-down control? Every built environment includes *some* deliberate design – especially in the public realm. Building façades are designed, roads are engineered, sidewalk widths are selected, and parks are laid out. Marshall (ibid.) suggests a critical role of urban design in delivering "functional complexity" for neighborhoods. Jacobs similarly argued for planners to generate diversity and supply what a neighborhood lacks. According to this stream of scholarship, urban design and planning can encourage diversity, adaptability, connectedness, resilience, and robustness – elements of healthy complex systems.

**Measuring Complexity**

Beyond qualitative formulations of complexity in urban form and design, how might it be measured and assessed? If cities are complex systems, and if complexity is important for systemic resilience, connectedness, and adaptability, then indicators of their complexity are essential for grounding debates and evaluating existing and proposed patterns and processes. However, no compilation or synthesis of relevant measures exists in the urban design literature to organize and connect these theories and methodologies to the practice of planning and designing resilient, connected, adaptable communities. While these measures are well-known to complexity scholars, many remain underexplored in design practice and research.

One stream of planning literature has considered quantitative measures of the urban form, but without explicitly engaging with complexity (e.g., Cervero and Kockelman 1997; Song and Knaap 2004; Tsai 2005; Clifton et al. 2008; Ewing and Cervero 2010; Schwarz 2010; Song et al. 2013b). Various complexity metrics at multiple scales, from metropolitan to neighborhood to building, are scattered throughout different bodies of literature. Lloyd (2001) surveyed and categorized measures of complexity across numerous fields of inquiry. Bourdic et al. (2012) provide an overview of cross-scale spatial indicators and touch on LEED-ND's neighborhood design assessment criteria. Additional surveys of complexity indicators for ecosystems and cities have been produced by Parrott (2010) and Salat et al. (2010).

The following discussion introduces, borrows, adapts, and reformulates relevant measures of complexity at the scale of urban design, touching on temporal measures but focusing on visual, spatial and structural measures applicable to urban morphology. In particular, it provides a quantitative framework organized around these key dimensions of complexity (Table 1) that accounts for both traditional urban design indicators as well as abstract measures arising from the complexity sciences. It reserves extended derivations and formulae for the references. This framework does not attempt to quantify all aspects of "good" design. Rather, it intends to formalize and measure the indistinct notion of *complexity* as it applies to urban design. Qualities related to vitality, sustainability, sense of place, and other important characteristics overlap, but are otherwise not the subject of this paper.

## Measures of Complexity

### Temporal Measures

As they only loosely relate to physical form/design, we briefly introduce temporal measures (which describe time series and system dynamics) before moving to form and structure. Temporal analysis techniques include embedding time series in higher dimensions, uncovering underlying attractors, estimating Lyapunov exponents, and analyzing the system from an information theoretic perspective (Batty 2005; Boeing 2016). Nonlinear analysis techniques from the physical sciences, such as reconstructing attractors or estimating Lyapunov exponents, have not been found to be particularly effective in the ecology literature (Parrott 2010).

Information theory, however, provides useful measures of complexity that may be applied to urban design. Shannon's (1948) theory of information entropy concerns the average amount of information contained in the revelation of a message or event. *Shannon entropy* indicates that the more types of things there are and the more equal each type's proportional abundance is, the less predictable the type of any single object will be (Boeing 2018b). This can be applied to abstract messages, time series, or spatial diversity (as discussed momentarily). Entropy is lowest when the system is highly ordered and thus completely predictable. It is highest when the system's disorder is maximized. Such a (category I) measure thus emphasizes disorder rather than peaking at some point between order and disorder (Batty 2005, Yeh and Li 2001).

Derived from Shannon entropy, *mean information gain* assesses how much new information is gained from each subsequent datum in a time series (Proulx and Parrott 2008) and *fluctuation complexity* measures the amount of structure within a time series by evaluating the order of and relationship between its values. In other words, how likely is it that we will observe some value *y* proximately after some other value *x*? Shannon entropy, mean information gain, and fluctuation complexity can be used to assess time series arising from urban systems. However, more usefully, they can be abstracted and re-appropriated to evaluate the human experience of moving through the physical space that results from urban design (Kuper 2017), which we turn to next.

**Visual Complexity**

In a simplified urban landscape, pedestrians gain little new information from the visual revelations of each passing step. However, highly complex (category I) urban environments bombard individuals with extensive new information as they move through space. In these examples, space is the medium and the unfolding visual tableaux are the message. This message could be discretized into arbitrary units such as meters, or into units relative to the specific urban landscape, such as street blocks or land parcels.

Much of the research on human perception of the built environment follows in the wake of Gibson's (1979) ecological framework and Appleton's (1975) prospect-refuge theory (e.g., Tveit et al. 2006; Ode et al. 2010). Clifton et al. (2008) discuss qualities of the urban form and human perceptions at multiple scales. For neighborhood and street scale urban design, perceptions of human scale are related to building heights and signage, perceptions of coherence are related to consistency of building heights, and sense of enclosure is related to building/element spacing and tree canopy. "Good" visual complexity tends to reach an optimum at some balance point between order and disorder, with "unity in variety," implying a (category II) convex interpretation (Elsheshtawy 1997; Gunawardena et al. 2015).

Ewing and Clemente (2013) perform a literature review yielding 51 perceptual qualities of urban environments, eight of which they select for further study because of their importance across the literature: imageability, enclosure, human scale, transparency, coherence, legibility, linkage, and visual complexity (cf. Ewing and Handy 2009). These researchers relate complexity to the number of perceptible differences a person is exposed to while moving through the city. Humans prefer to experience information at a comfortable rate – too little deprives the senses and too much overloads them. Good visual complexity depends on variety in buildings types, design details, street furniture, signage, human activity, sunlight patterns, and the rich textural details of street trees and urban forests. Complexity is lost when design becomes too top-down, controlled, and predictable in modern large-scale master plans. Poor complexity exists when urban design elements are too few, too similar and predictable, or too disordered to be comprehensible. In this formulation, complexity follows a category II convex function with a maximum value at some midpoint between order and disorder. Finally, Ewing and Clemente develop a field manual for measuring visual complexity as part of a larger toolkit for analyzing urban design according to their eight perceptual qualities. Cavalcante et al. (2014) provide an alternate, statistical image processing measure of urban visual complexity.

Fishman (2011) proposes that a significant conflict exists between the two primary paradigms of modern urban design. The first paradigm, spearheaded by the modernists, seeks to *open up* the dense and messy urban fabric with towers-in-the-park, spacing, highways, and technology. The second, espoused by neotraditionalists, seeks instead to *enclose space* through human-scale architecture, walkability, and a dense, complex, organic urban fabric. Jacobs and Appleyard (1987) argue that buildings in varied arrangements (in accordance with Fishman's second paradigm) enhance visual complexity, but interminable wide buildings – a hallmark of modernist design – detract from it (cf. Sussman and Hollander 2015). Jacobs (1995) further argues that buildings need multiple varied surfaces for light to move over to generate visual complexity. Macdonald (2005) explores how Vancouver generates visual complexity to put proverbial eyes on the street, with many entryways and interesting ground-level design.

Slow-moving pedestrians need a high level of complexity to hold their interest, but fast-moving motorists find that same environment chaotic. Dumbaugh and Li (2011) argue that urban designs that balance vehicle speeds, visual complexity, and traffic conflicts can increase motorist awareness, decrease collisions, and improve pedestrian safety. Marshall (2012b) contends that urban environments with perceptual richness are more interesting and enjoyable for humans, possibly because our species evolved in natural environments with high degrees of visual complexity.

**Spatial Measures**

Spatial complexity measures assess the system's patterns at snapshots in time rather than looking at dynamics over time. Shannon entropy can measure spatial complexity (Batty 2005) and mean information gain can measure ecosystem spatial complexity (Proulx and Parrott 2008). Boeing (2018b) adapts entropy measures to analyze urban spatial order through street orientations. Yeh and Li (2001) use entropy to monitor and track urban sprawl. Applying these information-theoretic metrics to space usually entails analyzing raster data for predictability.

Diversity and dispersion, however, are the most common spatial complexity measures in the urban design literature. Social diversity can enhance learning, adaptation, and unexpected social mixing. Jacobs (1961) praised diverse land uses for their ability to create synergies from complementary functions. Boarnet and Crane's (2001) behavioral framework of travel demand argues that urban design influences the (time) cost of travel by placing origins and destinations in closer proximity to (or further from) one another. Similarly, Cervero and Kockelman (1997) argue that land use diversity shapes travel behavior in urban environments. Salat et al. (2010) identify three types of urban spatial entropy related to complexity: diversity among similar objects, diversity in spatial distribution, and diversity of scale. Diversity among similar objects might refer to humans' income, ethnicity, employment, education, etc. It does, however, imply that even distributions score the highest, a questionable planning goal and reflection of complexity.

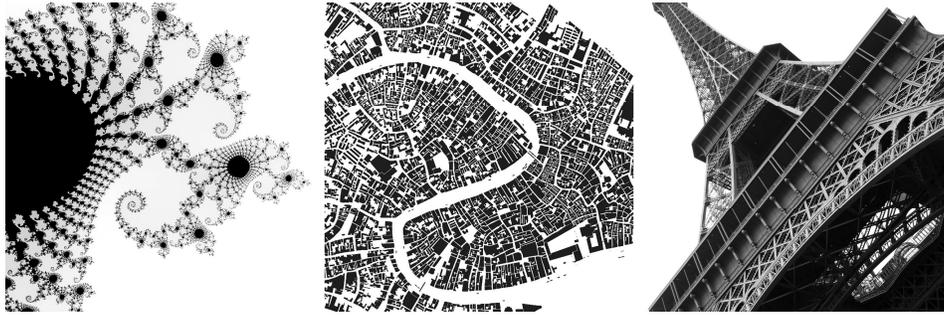

**Figure 2**. Left: the Mandelbrot set, a mathematical fractal. Center: Venice's fractal urban fabric. Right: the Eiffel Tower's fractal architecture. Image sources: author.

Wissen Hayek et al. (2015) measure land use mix and density to evaluate the quality of the neighborhood-scale urban environment. The Simpson diversity index measures the entropy of objects across space and is a common measure of land use mixture in urban design research (economists often call it the Herfindahl-Hirschmann index, while ecologists call it the probability of interspecies encounter). It is an *integral measure* that considers land use in a district as a whole, ignoring microscale structure (Song et al. 2013a). In contrast, a *divisional measure* reflects patterns within a district. Divisional measures better consider questions of scale. Dissimilarity indices measure how the land use mix within a district relates to the mix across the area as a whole (ibid.). Bordoloi et al. (2013) explore further measures of dissimilarity: these spatial distributions quantify how equitably some set of desirable or undesirable phenomena disperse across the city. For example, are schools clustered in wealthy neighborhoods rather than distributed evenly among all neighborhoods? Are waste treatment facilities concentrated in poor neighborhoods? Some clusters indicate spatial injustice, but other clusters can self-organize and emerge for inevitable or "good" reasons. Agglomeration economies cause job centers to cluster in certain areas (Jacobs 1969; Glaeser 2011) and the ecosystem services of urban forests are highest when green spaces are concentrated rather than evenly distributed throughout the built environment (Krasny et al. 2014).

Measures of topological structure assess the shape and physical configuration of a system. They are perhaps the most applicable measures of the complexity outcomes of urban design because they characterize that which is most dependent upon the design process: physical structure and topological arrangement. At the scale of urban design, structural measures fall primarily into two categories – scaling and connectivity – which we turn to next.

**Scaling**

Fractal structure refers to the roughness and self-similarity of some object, and how its detail relates to the scale at which it is observed (Jiang and Yin 2014). Fractals have similar structure at every scale rather than one single characteristic scale. In the real world, fractals do not perfectly exist at all spatial scales – from the infinitesimal to the infinite – as abstract mathematical fractals do (Figure 2). However, self-similarity of patterns over multiple scales exists throughout nature. Batty (e.g., 2005) has demonstrated how city structure and urban peripheries are fractal. Their features tend to be distributed according to a power law (a scale-free distribution) with few large items, a medium number of medium-sized items, and many small items.

Consider the example of an urban street network. At the largest scale, the city has a few major arterial roads and boulevards that serve as the key routes for system-wide traffic circulation. But if we zoom into this picture, a larger number of mid-sized collector streets appear, branching off from these few large arteries. As we zoom in further to a fine scale, a denser mesh of local streets appears, branching off from these collector streets. Certain distributions within a complex system may produce greater efficiency when they follow a power law rather than, say, an even distribution. For example, it is not ideal for a neighborhood to have the same number of arterial roads, collector streets, and local streets. Rather, there might be a small number of large arterial roads, a medium number of mid-sized collector streets, and a large number of capillary local streets. Such a system resembles a fractal. Murcio et al. (2015) measure urban transfer entropy to examine multi-scale patterns and flows. Similar fractal analyses can apply to the distribution and scaling of other urban structures such as buildings and land uses.

In particular, the fractal dimension measures how a form's complexity changes with regard to the scale at which it is measured. The fractal dimension of an object with one topological dimension refers to its space-filling characteristics that, through self-similarity, become a bit more than a one-dimensional line, yet a bit less than a two-dimensional plane. Measures of fractal dimension include the Hausdorff dimension and the box-counting dimension (Shen 2002). The concept of fractal dimensions can also be applied to two dimensional surfaces, such as the surface of a city, the surface of a building, or the surface of other urban design elements (Cooper et al. 2013). The Eiffel Tower, for example, exhibits fractal structure (Mandelbrot 1983). Fractals interweave qualities of scaling and visual complexity: while modernist architecture sought to erase complexity with simplified, segregated, sterile forms, both traditional architecture and today's paradigm tend to espouse organic forms with rich detail at multiple scales (Marshall 2008).

**Connectivity**

Network science provides a lens to explore structure through connectivity (Jiang 2016; Boeing 2017, 2018a; Turnbull et al. 2018). A network comprises a set of elements (called nodes) and their connections to one another (called edges). In a street network, nodes represent intersections and dead-ends, and edges represent the street segments that link them (Barthélemy 2011). Accessibility is a useful measure of urban design outcomes (and transportation/land use) related to network analysis. It concerns proximity, mobility, and social interaction within the public sphere. Popular "walkability" tools, such as WalkScore, and modeling tools such as UrbanSim use street networks to measure accessibility (Waddell et al. 2018). Urban networks can be measured for their complexity of structure particularly in terms of density, resilience, and connectedness (Figure 3), extending the toolkit commonly used by urban morphologists (Talen 2003; Marshall and Caliskan 2011).

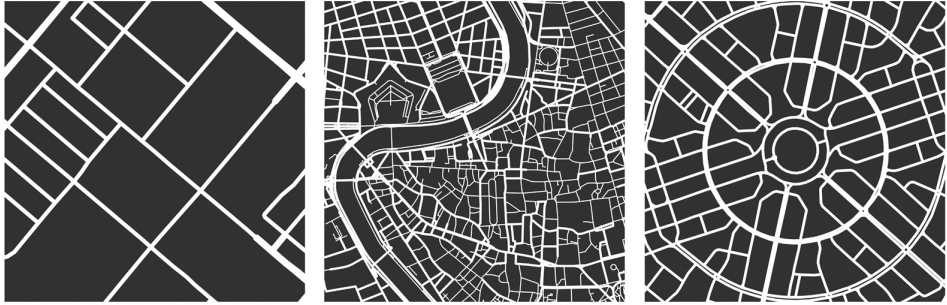

**Figure 3**. Street networks (one square mile each) exhibiting varying complexity through density, grain, connectivity, and permeability. Left: Irvine, California. Center: Rome, Italy. Right: Dubai, UAE. Image sources: author.

*Metric structure* refers to geometric areas, lengths, and familiar transportation and urban design variables (e.g., Cervero and Kockelman 1997; Masucci et al. 2009; Ewing and Cervero 2010). The *average street segment length* is a linear proxy for block size and indicates the network's grain. *Average circuity* measures the ratio of edge lengths to the great-circle distances between the nodes these edges connect, indicating the street pattern's efficiency (cf. Barthélemy 2011). These metrics, alongside node, intersection, edge, and street densities measure complexity by quantifying how the circulation network percolates through urban tissue, providing opportunities for movement and contact. However, connectivity *metrics* can behave inconsistently based on how study areas are drawn (Knight and Marshall 2015).

*Topological* measures of street network structure more robustly indicate connectedness and configuration. The network's *average node degree* topologically quantifies connectedness as the average number of edges incident to its nodes. The *average streets per node* adapts this for physical form rather than directed circulation, measuring the average number of physical streets that emanate from each node (i.e., intersection or dead-end). The *eccentricity* of a node is the longest shortest-path distance between it and every other node, representing how far it is from the node furthest from it (Urban and Keitt 2001). A network's *diameter* is its greatest eccentricity and its *radius* is its smallest eccentricity (Hage and Harary 1995). A network's *center* is the node with eccentricity equal to the radius. A network's *periphery* is the node with eccentricity equal to the diameter. These measures link fundamental complex network theory with analyses of urban design and mobility patterns. *Connectivity* represents the fewest number of nodes or edges that will disconnect the network if they are removed (Urban and Keitt 2001; O'Sullivan 2014) and thus indicates resilience. A network's *average node connectivity* more usefully denotes how many nodes must be removed on average to disconnect a randomly selected pair of nodes (Beineke et al. 2002). Brittle points of vulnerability characterize networks with low average connectivity: traffic jams and disruptions can arise when the urban form forces circulation through low-permeability choke points.

Clustering and centrality also indicate topological complexity, structure, and spatial distribution. A node's *clustering coefficient* represents the ratio between its neighbors' links and the maximum number of links that could exist between them (Opsahl and Panzarasa 2009). Jiang and Claramunt (2004) adapt this indicator to neighborhoods. Centrality specifies an element's importance in a network (Zhong et al. 2017). *Betweenness centrality* evaluates how many of the network's shortest paths pass through some node or edge (Barthélemy 2004; 2011). Barthélemy et al. (2013) operationalizes it to disambiguate the spatial signatures of Haussmann's renovation of Paris and the earlier self-organization and evolution of its urban fabric and street patterns. The maximum betweenness centrality measures the share of shortest paths that pass through the network's most important node: higher maximum betweenness centralities indicate networks more prone to inefficiency if this important choke point should fail. Other variations of centrality such as *closeness centrality* and *PageRank* can also be applied to street networks (Brin and Page 1998; Wang et al. 2011). Researchers often operationalize a basket of measures in unison to assess street network complexity. For instance, Porta et al. (2006) develop a *multiple centrality assessment* for urban street networks and use it to differentiate the signatures of planned cities versus self-organized cities. Crucitti et al. (2006) measure urban network centrality through a blend of closeness, betweenness, and information centralities.

Finally, we briefly consider space syntax theory. Thus far we have discussed street networks represented with their intersections as nodes and streets as edges. This is a *primal* representation. A *dual* representation inverts this topology, depicting the streets as nodes and the intersections as edges, and provides some analytical advantages in studying network structure (Crucitti et al. 2006). Space syntax studies this dual representation of urban street networks, analyzing axial streets and measuring the depth between edges (Hillier et al. 1976). However, space syntax's dual networks inherently disregard geometric, spatial, experiential information important to urban design that primal networks retain (Ratti 2004). Nevertheless, it underlies various adapted approaches to analytical urban design (Karimi 2012).

**Typology of Complexity Measures**

All of these methods of assessing the complexity of urban design, primarily at the neighborhood scale, can be organized into a preliminary typology (Table 2). The measures are grouped by Table 1's key dimensions: temporal, spatial, visual, scaling, and connectivity. While temporal measures assess the complexity of dynamics and process, the spatial, visual, and structural (i.e., scaling and connectivity) measures appear most promising for evaluating physical complexity at the scale of urban design.

| Dimension | Measure | Description |
|---|---|---|
| *temporal* | embedding time series | examine variables to reveal deep structure and patterns in data |
| *temporal, spatial* | Shannon entropy | how unpredictable a sequence is, based on number of types and proportional abundance |
| *temporal, spatial* | mean information gain | how much new information is gained from each subsequent datum |
| *temporal* | fluctuation complexity | amount of structure within a time series |
| *temporal, spatial* | urban transfer entropy | analytic tool for examining multi-scale urban patterns and flows |
| *visual* | Ewing and Clemente field guide | set of methods for assessing the physical, visual complexity of the streetscape |
| *visual* | Cavalcante streetscape measure | image processing method to assess visual complexity on contrast and spatial frequency |
| *spatial* | Simpson diversity index | assesses land use mix: how homogeneous or heterogeneous is the area of analysis |
| *spatial* | dissimilarity index | how does the land use mix within a subarea relate to the mix across the entire area |
| *scaling* | Hausdorff fractal dimension | how a form's complexity changes with regard to the scale at which it is measured |
| *scaling* | box-counting fractal dimension | how a form's complexity changes with regard to the scale at which it is measured |
| *spatial, connectivity* | destination accessibility | a function of land use entropy, amenity distribution, and network structure |
| *connectivity* | average streets per node | how well connected and permeable the physical form of the street network is, on average |
| *connectivity* | proportion of streets per node | characterizes the type, prevalence, and spatial distribution of intersection connectedness |
| *connectivity* | average street length | how long the average block is between intersections; proxy for areal block size and grain |
| *connectivity* | node/intersection, edge/street density | how fine- or coarse-grained the street network is |
| *connectivity* | average circuity | how similar network-constrained distances are to straight-line distances |
| *connectivity* | diameter/periphery, radius/center | network complexity in terms of max/min size, structure, and shape |
| *connectivity* | node/edge connectivity | what is the minimum number of elements that must fail to disconnect network? |
| *connectivity* | average node connectivity | average number of nodes that must fail to disconnect pair of non-adjacent nodes |
| *connectivity* | clustering coefficient | extent to which the neighbors of some node are connected to each other |
| *connectivity* | average clustering coefficient | mean of the clustering coefficients for all nodes |
| *connectivity* | betweenness centrality | the importance of an element in terms of how many shortest paths pass through it |
| *connectivity* | average betweenness centrality | mean of the betweenness centralities for all nodes |
| *connectivity* | closeness centrality | elements rank as more central if they are on average closer to all other elements |
| *connectivity* | average closeness centrality | mean of the closeness centralities for all elements |
| *connectivity* | PageRank | ranking of node importance based on structure of incoming links |
| *connectivity* | multiple centrality assessment | uses primal, metric networks to examine multiple indices of centrality |
| *connectivity* | space syntax | uses dual, topological networks to examine closeness centrality of a named street |

**Table 2.** Typology of urban form/design complexity measures.

## Discussion

Cities are complex systems composed of many human agents interacting in physical urban space. Urban design has evolved through eras of classicism, organicism, austere modernism, postmodernism, and neotraditionalism – each of which encounters the city's physical complexity with different goals. Multiple synergies exist today between urban design objectives and recent knowledge emerging from the complexity sciences as complexity theory. Complexity underlies urban resilience, accessibility, and livability. Path dependence, hysteresis, and historical accidents all arise in complex systems and shape the future of the urban form (Siodla 2015). Complexity makes urban environments more resilient and robust, providing greater opportunities for social encounter, mixing, and adaptation through social learning. Complexity entails greater connectivity, diversity, variety, and sustainability. Today, prominent urban design movements such as the new urbanism and smart growth openly embrace complexity (Duany, Plater-Zyberk & Co. 2001; Talen 2003; Sanders 2008; Congress for the New Urbanism 2015). But to better foster the benefits of healthy complex adaptive systems, urban design must move beyond metaphor and inspiration to more precisely formalize what complexity is and how we can measure it to assess both the world as it is, and proposals for how it could be instead.

    This paper takes a step in this direction by compiling a toolkit of indicators and measures of complexity to analyze resilience, connectedness, and adaptive capacity at the scale of urban design. First it introduced three interpretative categories of complexity (Figure 1), before honing in on how category II – the structured balance between order and chaos – pertains to urban form. Next it identified five key dimensions of complexity: temporal, visual, spatial, scaling, and connectivity (Table 1). It then presented various measures, indicators, and examples of each dimension and how they relate to urban form and design. Finally, it collated these into a typology to assess urban form, design claims, and project outcomes (Table 2). The analytical framework developed here is generalizable to empirical research of various neighborhood types and design standards. This typology draws from multiple scientific disciplines to identify how an urban system's physical form organizes complex human interactions and connections – thus linking structure and dynamics – particularly at urban design's scale of intervention.

    How does this knowledge lead to different or better urban design? First, the categories can help urban design theory and practice critically evaluate and normatively balance complexity goals based on local culture and politics. Urban design, as a channel of human self-organization, can produce various balances of order and variety in its streetscapes (anywhere from sterile monotony to bewildering overstimulation), circulation networks (from compact grids, to sprawling loops-and-lollipops, to dense meshes of interwoven paths), land uses (from single-use functional zoning to intermixed variety), and social character (from segregation and disconnection to integration, contact, and exchange).

    Second, to critically evaluate these possible futures beyond metaphor and inspiration, the five dimensions of complexity and their attendant measures can ground debate and analysis. They organize and quantify characteristics of current and proposed urban design that shape circulation, contact, access, choice, resilience, adaptability, and

equity. For example, this provides new ways to evaluate how thoroughly linked and permeable a network is, and how it can be resilient against floods, earthquakes, traffic collisions, congestion, and other disruptions. It grounds design decisions in science and theory to clarify goals and evaluate the outcomes of alternative plans. These measures help designers evaluate how well the physical built environment might adapt to expected – or unexpected – change. They offer a rubric for the continuing shift of design theory and practice away from a logic of top-down artificially constructed urban landscapes and towards a logic of organic growth, evolution, and resilience. If cities are complex systems, then a key goal of urban design must be to support their complex functioning by producing a physical substrate conducive to the values of resilience, connectedness, adaptability, and equity. This paper synthesized a set of key measures to evaluate these forms, goals, and actions rigorously.